# A Model of Neutrino Masses and Mixing from Hierarchy and Symmetry

Peter Kaus[a], Sydney Meshkov[b]

[a] Physics Department, University of California, Riverside, CA 92521, USA
[b] California Institute of Technology, Pasadena, CA 91125, USA

**Abstract**

We construct a model that allows us to determine the three neutrino masses directly from the experimental mass squared differences, $\Delta_{atm}$ and $\Delta_{sol}$, together with the assumption that $\Lambda=\sqrt{(1/6)} \equiv \sqrt{(m_2/m_3)}$. The parameter, $\Lambda$, basically a Clebsch-Gordan coefficient with the value of about 0.4 stems from the group $S_3$, and is NOT an expansion parameter, in contrast with the Wolfenstein parameter, $0.22 < \lambda < 0.25$ needed to explain quark masses. For a variety of initial values of $\Delta_{atm}$, we find that the lowest mass, $m_1$, varies from $1.4 - 3.6 \times 10^{-3}$ eV, the next lowest mass, $m_2$, varies only slightly from $8.4 - 9.0 \times 10^{-3}$ eV, and the heaviest mass, $m_3$, ranges from $5.0 - 5.4 \times 10^{-2}$ eV. The elements of the mixing matrix, U, and of the mass matrix, M, are examined with particular emphasis on the role of small angle $\theta_{13}$. The phase, $\delta$, of the mixing matrix U has a serious effect on the mass matrix only for the matrix elements $M_{e\mu}$ and $M_{e\tau}$, because these are the only ones for which the real part vanishes in the allowed range for $\theta_{13}$. Their dependence on $s_{13}$ for various values of $\delta$ is given explicitly. We study the elements of the mass matrix, M, for our solution 1, that of the perfect rational hierarchy, for the case $\delta = 0$, and find that all of them are smaller than 0.03 eV. The only candidates for double texture zero models are $M_{ee}$ and $M_{e\mu}=M_{\mu e}$.

**1. Introduction**

In the Standard Model, the twelve masses of the three generations of four families are arbitrary. Unification of quarks and leptons will eliminate these capricious numbers or at least establish strong relations between them. For a very promising approach see [1].

At the present time, one thing that we can do is to look for patterns. In particular, the 'rational' hierarchy of quarks and charged leptons is well confirmed. By rational here we mean that mass-ratios of members of a family are very close to powers of a parameter $\lambda$ [2]. For example, $m_b:m_s:m_d \approx 1:\lambda^2:\lambda^4$. Furthermore, this parameter dominates the symmetry breaking exhibited by the mixing angles of the unitary matrix, which gives the flavor states as linear combinations of the mass eigenstates.

Mass patterns for neutrinos are quite different. The information for neutrinos comes mainly from solar and atmospheric neutrino oscillations [3], [4].

$$\Delta_{sol} = |m^2_{\nu 2} - m^2_{\nu 1}| \approx 6.9 \times 10^{-5} \text{ eV}^2 \quad \text{and} \quad \Delta_{atm} = |m^2_{\nu 3} - m^2_{\nu 2}| \approx 2.6 \times 10^{-3} \text{ eV}^2$$

In the following we determine the neutrino masses by proposing a new relation between the mixing angles and the mass ratios. The mixing angle $\theta_{13}$ is small. In the limit $\theta_{13}$ goes to zero, we impose S3-S2 symmetry on the mixing matrix to fix the remaining mixing angles $\theta_{23}$ and $\theta_{12}$. In a strong hierarchical model ($m_1$ small), we must have $m_2 \approx \sqrt{\Delta_{sol}}$ and $m_3 \approx \sqrt{(\Delta_{atm} + \Delta_{sol})}$. Motivated in part by the observed numerical similarity of $s_{12}s_{23}$ and $(\Delta_{sol}/\Delta_{atm})^{(1/4)}$, we equate the Cabibbo angle $\sqrt{(m_2/m_3)}$ to $s_{12}s_{23}$, which will be named

Λ, similar in spirit, but not in magnitude to the Wolfenstein parameter λ. With this identification, the masses and the mass matrix are totally determined by $\Delta_{sol}$ and $\Delta_{atm}$.

## 2. Symmetry and Hierarchy Lead to a Proposed New Parameter for Neutrino Mass Determination

The flavor states $\nu_e, \nu_\mu$ and $\nu_\tau$ are related to the mass eigenstates $\nu_1$, $\nu_2$ and $\nu_3$ by the unitary transformation U.

$$U = \begin{pmatrix} c_{12}c_{13} & -s_{12}c_{13} & s_{13}e^{-i\partial} \\ s_{12}c_{23} + c_{12}s_{13}s_{23}e^{i\partial} & c_{12}c_{23} - s_{12}s_{13}s_{23}e^{i\partial} & -c_{13}s_{23} \\ s_{12}s_{23} - c_{12}s_{13}c_{23}e^{i\partial} & c_{12}s_{23} + s_{12}s_{13}c_{23}e^{i\partial} & c_{13}c_{23} \end{pmatrix} \quad (1)$$

There are two 'large' angles $\theta_{12}$ and $\theta_{23}$. Setting the small angle $\theta_{13}$, for which there is as yet no lower limit, equal to zero, we obtain $U_o$:

$$U_0 = \begin{pmatrix} c_{12} & -s_{12} & 0 \\ s_{12}c_{23} & c_{12}c_{23} & -s_{23} \\ s_{12}s_{23} & c_{12}s_{23} & c_{23} \end{pmatrix} \quad (2)$$

The three columns of U are the three eigenvectors of the mass matrix in the $\theta_{13}=0$ limit. If $V_i$ is the ith column of U (i=1,2,3), then the mass matrix M is given by:

$$M = \sum_i m_i V_i V_i^\dagger \quad (3)$$

where $m_i$ is the ith eigenvalue of M.

It has been proposed more than 15 years ago that the 'mass gap' of the hierarchical pattern is associated with pairing forces in analogy with Cooper pairs in BCS theory and the mass matrix of the neutral pseudoscalar mesons [5]. In this limit, the mass matrix is 'democratic' [6] and when diagonalized gives rise to only one massive state, the coherent state. The 'democratic' vector $V_d$ is of particular interest here, where

$$V_d = \sqrt{(1/3)} \begin{bmatrix} 1 \\ 1 \\ 1 \end{bmatrix} \quad (4)$$

and

$$V_d V_d^\dagger = (1/3) \begin{pmatrix} 1 & 1 & 1 \\ 1 & 1 & 1 \\ 1 & 1 & 1 \end{pmatrix}, \quad (5)$$

the 'democratic' matrix. The vector $V_d$ was assigned to the heaviest mass, $m_3$, with pairing forces creating the mass gap in mind. The masses $m_2$ and $m_1$ were thought to be generated through a breaking of this $S_3$ symmetry, $S_3 \to S_2 \to S_1$ [5], [7].

However, the smallness (or vanishing) of $\theta_{13}$ makes the BCS type mass gap interpretation for $m_3$ untenable in the neutrino case. In contrast to the BCS case, because of the consequent vanishing of $U_{e3}$ in $U_o$ (Eq. (2)), we have $m_3$ as a coherent mixture of $m_{\nu\mu}$ and $m_{\nu\tau}$, which agrees with maximal mixing, $\theta_{23} = \pi/4$. We now have $S_2$ symmetry for $m_3$ and reserve $S_3$ symmetry for $m_2$. This, in fact, completely determines $U_0$. This assignment of $V_d$ as the eigenvector for $m_2$ has lately received considerable attention in the literature [8]. $S_2$ symmetry for $V_3$ implies $\theta_{23} = \pi/4$ (maximal mixing) and $s_{23} = c_{23} = \sqrt{(1/2)}$.

We now relate the second large mixing angle, $\theta_{12}$, to the mass ratio $m_2/m_3$ by the relation:

$$-s_{12}\, s_{23} = \sqrt{(m_2/m_3)} \equiv \Lambda. \tag{6}$$

This association of the mixing angles with the mass ratios was suggested by us earlier on phenomenological grounds [7], because both $s_{12}s_{23}$ and $(\Delta_{sol}/\Delta_{atm})^{(1/4)}$ are about the same, approximately equal to 0.4. We propose it here as a `natural' pattern.

Considering $s_{12}$ a small parameter for the moment (it is not), we get to first order in $s_{12}$ ($c_{12}=1$) the matrix $u_0$:

$$u_0 = \begin{pmatrix} 1 & \sqrt{2}\,\Lambda & 0 \\ -\Lambda & \sqrt{1/2} & -\sqrt{1/2} \\ -\Lambda & \sqrt{1/2} & \sqrt{1/2} \end{pmatrix} \tag{7a}$$

or more suggestively:

$$u_0 = \begin{pmatrix} 1 & \Lambda & \Lambda \\ -\Lambda & 1 & 0 \\ -\Lambda & 0 & 1 \end{pmatrix} \begin{pmatrix} 1 & 0 & 0 \\ 0 & \sqrt{1/2} & -\sqrt{1/2} \\ 0 & \sqrt{1/2} & \sqrt{1/2} \end{pmatrix} \tag{7b}$$

This shows the dynamic role assigned to $\theta_{12}$ by the assumption (6) and why we may consider it as 'natural'.

Restoring $c_{12}$ and full unitarity we have for $U_0$:

$$U_0 = \begin{pmatrix} \sqrt{(1-2\Lambda^2)} & \sqrt{2}\,\Lambda & 0 \\ -\Lambda & \sqrt{1/2}\sqrt{(1-2\Lambda^2)} & -\sqrt{1/2} \\ -\Lambda & \sqrt{1/2}\sqrt{(1-2\Lambda^2)} & \sqrt{1/2} \end{pmatrix} \quad (8)$$

Imposing $S_3$ symmetry (democracy) for the vector $V_2$ implies $U_{e2}=U_{\mu 2}=U_{\tau 2}$ or $\sqrt{2}\,\Lambda = \sqrt{(1/2)}\sqrt{(1-2\Lambda^2)}$, so that

$$U_0 = \begin{pmatrix} 2\Lambda & \sqrt{2}\,\Lambda & 0 \\ -\Lambda & \sqrt{2}\,\Lambda & -\sqrt{1/2} \\ -\Lambda & \sqrt{2}\,\Lambda & \sqrt{1/2} \end{pmatrix} \quad (9)$$

By normalization, it follows that

$$\sqrt{(m_2/m_3)} \equiv \Lambda = \sqrt{(1/6)} \quad (10)$$

Of course $\sqrt{(1/6)}$ is not a capricious number, along with $\sqrt{(1/2)}$ it is a Clebsch-Gordan Coefficient, but that it should be equal to $\sqrt{(m_2/m_3)}$ is a capricious notion.

In a hierarchical model, with small or vanishing $m_1$, we have

$$m_2/m_3 = \sqrt{(\Delta_{sol} + m_1^2)}/\sqrt{(\Delta_{atm} + \Delta_{sol} + m_1^2)} \approx \sqrt{(\Delta_{sol}/\Delta_{atm})}.$$

It is, of course entirely possible that it is a coincidence that $s_{12}s_{23} \approx (\Delta_{sol}/\Delta_{atm})^{(1/4)}$ and that both are approximately equal to $\sqrt{(1/6)}$, which is the value demanded by S3-S2 symmetry, but we make it the basis of the present model. Hence Eq (10).

We now have $\sin(\theta_{23}) = \cos(\theta_{23}) = \sqrt{(1/2)}$ and $\sin(\theta_{12}) = \sqrt{(1/3)} = \sqrt{2}\,\Lambda$, so that:

$$\tan^2(\theta_{23}) = 1 \qquad \tan^2(\theta_{12}) = 1/2 \quad (11)$$

The hierarchy indicated here is not very strong, $m_2 = \Lambda^2 m_3 = (1/6) m_3$, and $\Lambda$ should not be used as an expansion parameter. In fact, the situation is very different from the quark sectors. There, the possible $S_3$-$S_2$ symmetry is presumably the same for the d and u sectors and does not appear in the $V_{ckm} = U_d^\dagger U_u$, which is then just 1. Only the symmetry breaking terms, dominated by powers of $\lambda \approx 0.23$ are seen and the underlying symmetry, if it exists, is obscured in the resulting Wolfenstein representation. The mixing angles can be large or small, depending on the assumed flavor basis. In the present model, on the other hand, $\Lambda$ is intrinsic to the symmetry and must be $\sqrt{(1/6)} \approx 0.4$.

## 3. The neutrino mass spectrum

Assuming the normal ordering of masses, $m_1^2 < m_2^2 < m_3^2$ we have two equations, with the mass squared differences, $\Delta_{sol}$ and $\Delta_{atm}$ determined:

$$m_2^2 = \Delta_{sol} + m_1^2 \quad (12a)$$
$$m_3^2 = \Delta_{atm} + \Delta_{sol} + m_1^2 \quad (12b)$$

But now a mass scale is provided by the hypothetical third equation (6), relating the two masses

$$\sqrt{(m_2/m_3)} \equiv \Lambda = \sqrt{(1/6)},$$

or

$$\sqrt{(\Delta_{sol} + m_1^2)}/\sqrt{(\Delta_{atm} + \Delta_{sol} + m_1^2)} = (1/6) \quad (13)$$

Clearly, a real solution for $m_1$ is possible only if $\sqrt{\Delta_{sol}}/\sqrt{(\Delta_{atm} + \Delta_{sol})} \leq (1/6)$.

Using, for the moment, $\Delta_{sol} \approx 6.9 \times 10^{-5}$ eV$^2$ and $\Delta_{atm} \approx 2.6 \times 10^{-3}$ eV$^2$, gives

$$\sqrt{\Delta_{sol}}/\sqrt{(\Delta_{atm} + \Delta_{sol})} \approx 1/6.2 \quad (14)$$

which means, that whatever values we take for $\Delta_{sol}$ and $\Delta_{atm}$, consistent with the data, $m_1$ will certainly be small.

Without loss of generality, but with an eye towards 'rational' hierarchy, we now represent the masses $m_1$, $m_2$, $m_3$ in terms of parameters $\Lambda$, $\rho$ and $m_3$:

$$M_{diag} = \begin{pmatrix} m_1 & 0 & 0 \\ 0 & \sqrt{\Delta_{sol} + m_1^2} & 0 \\ 0 & 0 & \sqrt{\Delta_{atm} + \Delta_{sol} + m_1^2} \end{pmatrix} = m_3 \begin{pmatrix} \rho\Lambda^4 & 0 & 0 \\ 0 & \Lambda^2 & 0 \\ 0 & 0 & 1 \end{pmatrix} \quad (15)$$

Thus $m_2/m_3 = \Lambda^2$ and $m_1/m_3 = \rho \Lambda^4$ and $\Lambda = \sqrt{(1/6)}$. 'Perfect' rational hierarchy would mean $\rho = 1$. The data for $\Delta_{sol}$ and $\Delta_{atm}$ considerably restrict the possible solutions.

In Fig. 1 we use

$$\Delta_{atm} = \Delta_{sol}\left((\Lambda^4 - 1)/\Lambda^4(\Lambda^4 \rho^2 - 1)\right), \quad (16)$$

which follows from the representation (15), to see the range, if any, of solutions consistent with the experimental range of $\Delta_{sol}$ and $\Delta_{atm}$. For clarity of the figure we chose a rectangle slightly smaller than the 2$\sigma$ limits of M. Maltoni et.al., and of Ishitsuka [3], [4],

$$6.0 \ 10^{-5} \text{ eV}^2 < \Delta_{sol} < 8.4 \ 10^{-5} \text{ and } 1.8 \ 10^{-3} < \Delta_{atm} < 3.3 \ 10^{-3}.$$

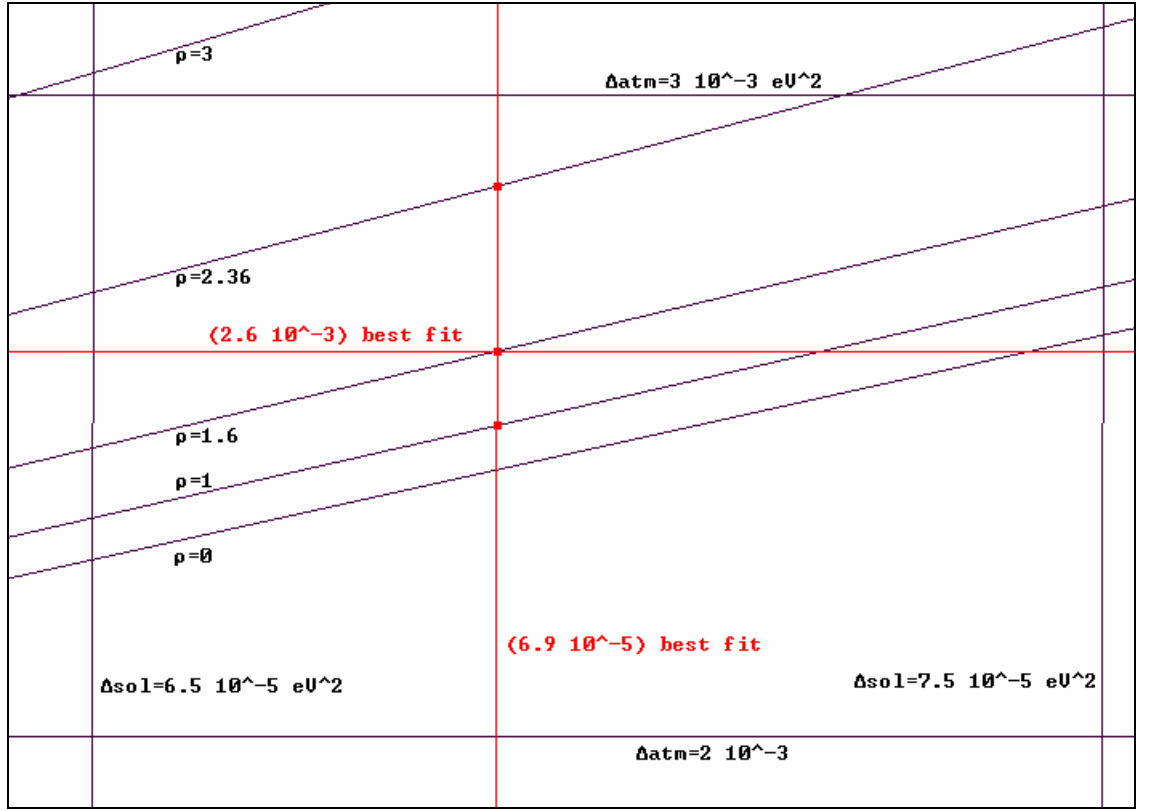

*Fig.1 $\Delta_{atm}$ vs $\Delta_{sol}$ (Eq. 16 with $\Lambda^2=1/6$) for various values of $\rho$. ($m_1/m_2=\rho\, m_2/m_3=\rho\Lambda^2$).
Acceptable solutions are within the (slightly arbitrary) rectangle $2\,10^{-3}\, eV^2 < \Delta_{atm} < 3\,10^{-3}\, eV^2$
and $6.5\,10^{-5}\, eV^2 < \Delta_{sol} < 7.5\,10^{-5}\, eV^2$. Three solutions are marked. They correspond to
$\Delta_{sol} = 6.9\,10^{-5}\, eV^2$, with $\rho=1$ (rational hierarchy), $\rho=1.6$ (best fit),[3] and $\rho=2.36$ (double texture
zero, see text) respectively.*

With $\Delta_{sol}$ and $\Delta_{atm}$ given, all mass values are fixed. It is gratifying to have $\rho \approx 1$ to be squarely in the data range, because values of $\rho$ much different from unity stress the spirit of rational hierarchy. It is clear from Fig.1 that $\rho$ can also be smaller than 1, even zero, without deviating from the 'best fit' by much. We now give details for three solutions.

Equations (15) can be solved to give the masses and the ratio parameter $\rho$, ($m_1/m_2=\rho\, m_2/m_3=\rho\Lambda^2$), as functions of $\Delta_{atm}$ and $\Delta_{sol}$.

$$m_1^2 = \Delta_{atm} \frac{\Lambda^4}{(1-\Lambda^4)} - \Delta_{sol} \qquad (17a)$$

$$m_2^2 = \Delta_{atm} \frac{\Lambda^4}{(1-\Lambda^4)} \qquad (17b)$$

$$m_3^2 = \Delta_{atm} \frac{1}{(1-\Lambda^4)} \qquad (17c)$$

$$\rho^2 = (1/\Lambda^4)((\Delta_{atm}+\Delta_{sol})/\Delta_{atm}) - (1/\Lambda^8)(\Delta_{sol}/\Delta_{atm}) \qquad (17d)$$

**6**

The equations (17) are valid independent of the choice of Λ. Substituting Λ=√(1/6) from Eq. (10), the three special solutions marked in Fig,1 are then determined.

Input: $\Lambda^2=1/6$, $\Delta_{sol}=6.9 \cdot 10^{-5}$ eV$^2$ and ρ.

SOLUTION-1 'perfect' rational hierarchy  ρ=1

$\Delta_{atm}= 2.5 \cdot 10^{-3}$ eV$^2$
$m_1 = 1.4 \cdot 10^{-3}$ eV
$m_2 = 8.4 \cdot 10^{-3}$ eV
$m_3 = 5.0 \cdot 10^{-2}$ eV

SOLUTION-2   best fit   ρ= 1.6

$\Delta_{atm} = 2.6 \cdot 10^{-3}$ eV$^2$

$m_1 = 2.3 \cdot 10^{-3}$ eV
$m_2 = 8.6 \cdot 10^{-3}$ eV
$m_3 = 5.2 \cdot 10^{-2}$ eV

SOLUTION-3   Double Texture-0   (with $\sin(\theta_{13})=0.11$)   ρ=2.36

$\Delta_{atm}= 2.9 \cdot 10^{-3}$ eV$^2$
$m_1 = 3.6 \cdot 10^{-3}$ eV
$m_2 = 9.0 \cdot 10^{-3}$ eV
$m_3 = 5.4 \cdot 10^{-2}$ eV

All three solutions have the property $m_3= 6 m_2$ and $\rho m_2 = 6 m_1$ with $0<\rho<3$. $\Delta_{atm} = |m_3^2 - m_2^2|$ and $\Delta_{sol} = |m_2^2 - m_1^2|$ are within the acceptable experimental limits. All masses listed are absolute values.

## 4. Elements of the Mass Matrix and their Properties

The Mass matrix M is given by

$$M = U M_d U^\dagger \tag{18}$$

where U is given by (1) and

$$M_d = \begin{pmatrix} m_1 & 0 & 0 \\ 0 & m_2 & 0 \\ 0 & 0 & m_3 \end{pmatrix} \tag{19}$$

The nine elements of M are then given by:

$$M_{ee} = (m_1 + s_{12}^2 (m_2 - m_1)) - s_{13}^2 (m_1 - m_3 + s_{12}^2 (m_2 - m_1)) \tag{20a}$$

$$M_{e\mu} = M_{\mu e}^* = \sqrt{(1/2)}\, c_{13} (c_{12} s_{12} (m_1 - m_2) + s_{13} e^{-i\delta} (s_{12}^2 (m_2 - m_1) + m_1 - m_3)) \tag{20b}$$

$$M_{e\tau} = M_{\tau e}^* = \sqrt{(1/2)}\, c_{13} (c_{12} s_{12} (m_1 - m_2) + s_{13} e^{-i\delta} (s_{12}^2 (m_1 - m_2) - m_1 + m_3)) \tag{20c}$$

$$M_{\mu\mu} = s_{12} s_{13} c_{12} (m_1 - m_2) \cos(\delta) + (1/2)( c_{13}^2 s_{12}^2 (m_1 - m_2) + c_{13}^2 m_3 + s_{13}^2 m_1 + m_2) \tag{20d}$$

$$M_{\mu\tau} = M_{\tau\mu}^* = -(1/2)((m_1 - m_2)(c_{12}^2 s_{13}^2 - s_{12}^2) + (m_3 - m_2) c_{13}^2) - i\, c_{12} s_{12} s_{13} (m_2 - m_1) \sin(\delta) \tag{20e}$$

$$M_{\tau\tau} = c_{12} s_{12} s_{13} (m_2 - m_1) \cos(\delta) + (1/2)( s_{12}^2 c_{13}^2 (m_1 - m_2) + c_{13}^2 m_3 + s_{13}^2 m_1 + m_2). \tag{20f}$$

The matrix elements, Eqs. (20), are the same whether the neutrinos are Dirac or Majorana particles. This may be seen by writing $U_{maj}$ as $U_{maj} = U \times \mathrm{diag}(e^{i\alpha_1/2}, e^{i\alpha_2/2}, 1)$ and evaluating $M_{maj}$ using Eqs. (18) and (19). The phases $\alpha_1$ and $\alpha_2$ are the Majorana phases. While these phases do not affect the mass matrix, they do have physical consequences for neutrinoless double beta decay.

We now explore some of the features of M. The angles of the mixing matrix U are:

$$\sin(\theta_{12}) = -\sqrt{(1/3)} \qquad \tan^2(\theta_{12}) = (1/2)$$
$$\sin(\theta_{23}) = \sqrt{(1/2)} \qquad \tan^2(\theta_{23}) = 1$$

For all the examples we have chosen the solutions (1,2,3) given above, so that $\Delta_{sol} = 6.9\, 10^{-5}$ eV$^2$ with $\Delta_{atm} = (2.5, 2.6, 2.9)\, 10^{-3}$ eV$^2$ respectively. We have also chosen $m_2 = -|m_2|$.

Fig.2 shows the elements of M for Solution-1 as functions of $\sin(\theta_{13})$ with $\delta=0$. The maximum allowed $\sin(\theta_{13}) \approx 0.25$.

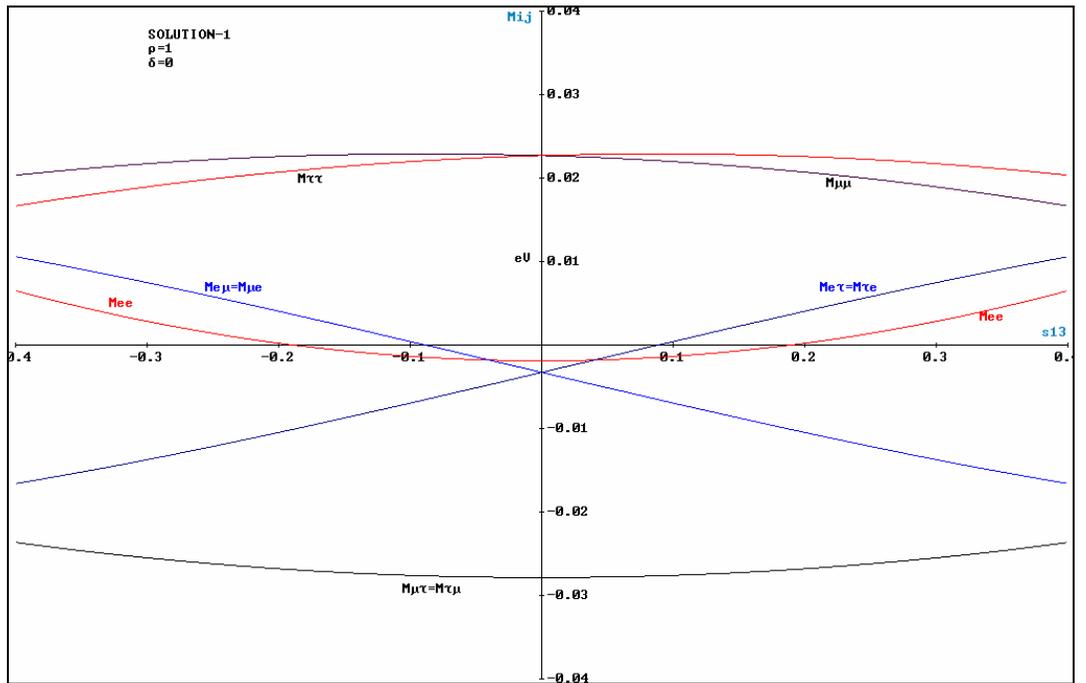

*Fig.2 Elements of the Mass matrix M as functions of sin($\theta_{13}$), with $\delta$=0. The masses are from Solution-1, $\rho$=1, the rational hierarchy solution, $\Delta_{sol}$=6.9 $10^{-5}$ eV$^2$ and $\Delta_{atm}$=2.5 $10^{-3}$ eV$^2$, and $m_2$= - 8.4 $10^{-3}$eV. All elements are smaller than 0.03 eV.*

For making theories, the question of double texture zeroes [9] is of some interest. From Fig.2 it is clear that $M_{ee}$ and $M_{e\mu}=M_{\mu e}$ are the only good candidates. (Or $M_{e\tau}=M_{\tau e}$ with the opposite sign of $\theta_{13}$).

In Fig.3 we compare $M_{ee}$, $M_{e\mu}$ and $M_{e\tau}$ for Solution-1 ($\rho=1$) and Solution-3 ($\rho=2.36$). This clearly shows that $M_{ee}=M_{e\mu}=0$ for $\rho=2.36$ and $\sin(\theta_{13}) \approx -0.11$.

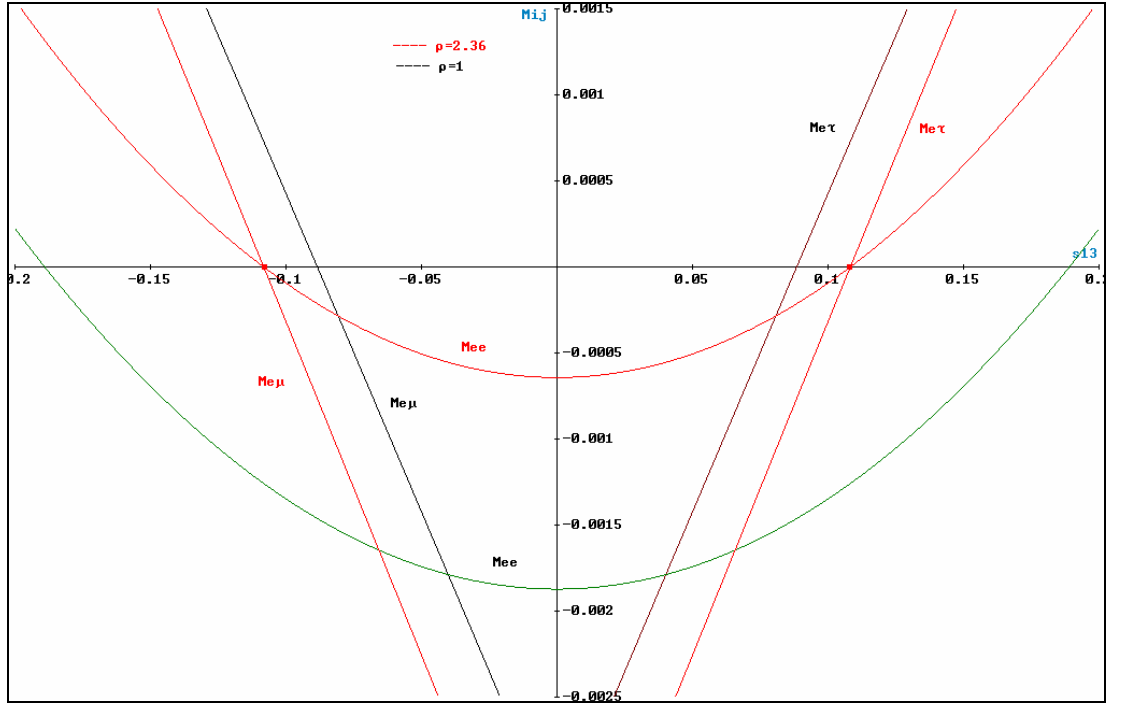

*Fig.3 $M_{ee}$, $M_{e\mu}$ and $M_{e\tau}$. $\Delta_{sol}=6.9\ 10^{-5}\ eV^2$ with $\rho=1$, $\Delta_{atm}=2.5\ 10^{-3}\ eV^2$ and $\rho=2.36$, $\Delta_{atm}=2.9\ 10^{-3} eV^2$. The double texture zeros for $\sin(\theta_{13})= \pm 0.11$ are marked.*

The mass matrix for the $\rho=2.36$ solution, with $\sin(\theta_{13})=-0.11$, is given by

$$M = (5.4\ 10^{-2}) \begin{pmatrix} 0 & 0 & -0.15 \\ 0 & 0.46 & -0.54 \\ -0.15 & -0.54 & 0.44 \end{pmatrix} eV \qquad (21)$$

Even though we said earlier that $\Lambda$ is not an expansion parameter, if we expand to leading orders of $\Lambda$, we find that the double texture zero solution (Solution-3, with $\sin(\theta_{13}) \approx -0.11$) is well approximated by:

$$M = m_3 \begin{pmatrix} 0 & 0 & -6\Lambda^4 \\ 0 & 1/2-(3/2)\Lambda^4 & -1/2-(3/2)\Lambda^4 \\ -6\Lambda^4 & -1/2-(3/2)\Lambda^4 & 1/2-(3/2)\Lambda^4 \end{pmatrix} \qquad (22)$$

with $\sin(\theta_{13}) \approx -3\sqrt{2}\ \Lambda^4$ and $\rho \approx 1/2\Lambda^2$ to lowest order.

The phase δ of the mixing matrix U has a serious effect for the mass matrix only for the matrix elements $M_{e\mu}$ and $M_{e\tau}$, because these are the only ones for which the real part vanishes in the allowed range for $\theta_{13}$.

Fig.4 shows $|M_{e\mu}|$ vs. $\sin(\theta_{13})$ for various values of δ. The values for $|M_{\mu e}|$ are identical as well as for $|M_{e\tau}|$ and $|M_{\tau e}|$, with $\theta_{13} \to -\theta_{13}$.

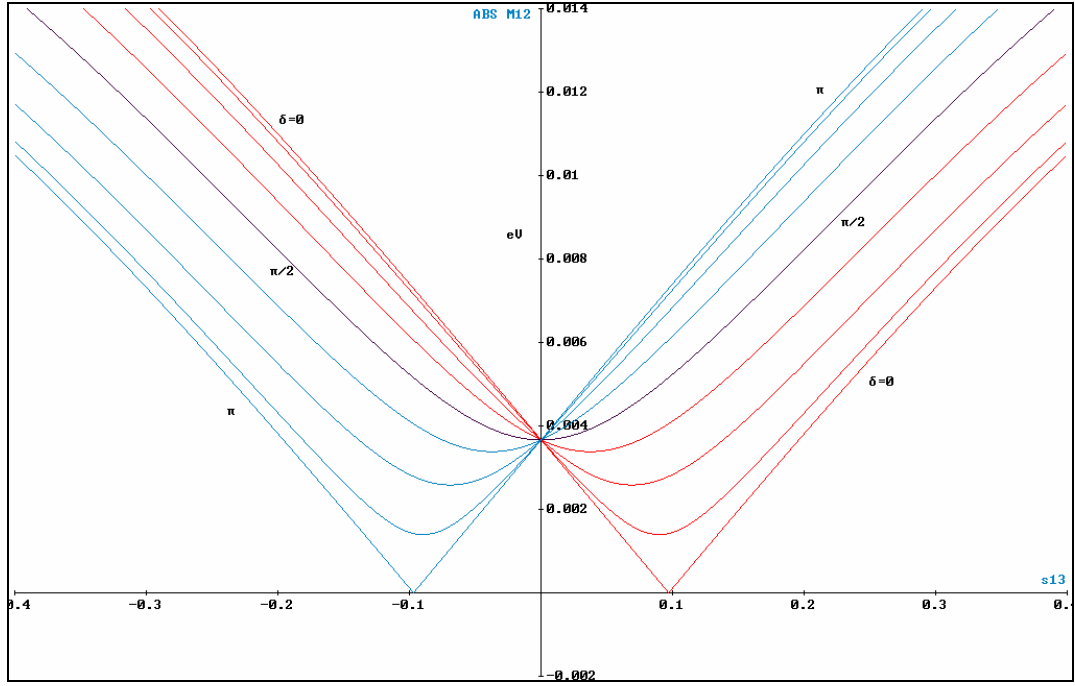

*Fig. 4 $|M_{e\mu}|$ vs. $\sin(\theta_{13})$ for various values of δ, $0 \leq \delta \leq \pi$ in steps of $\pi/8$ for Solution 2.*

**CONCLUSIONS**

We have shown that the neutrino masses may be evaluated directly from the experimental mass squared differences, $\Delta_{atm}$ and $\Delta_{sol}$ together with the assumption that $\Lambda = \sqrt{(1/6)} \equiv \sqrt{(m_2/m_3)}$.
For a variety of initial values of $\Delta_{atm}$ the lowest mass, $m_1$, varies from $1.4 - 3.6 \; 10^{-3}$ eV,
The next lowest mass, $m_2$, varies only slightly from $8.4 - 9.0 \; 10^{-3}$ eV, and the heaviest mass, $m_3$, ranges from $5.0 - 5.4 \; 10^{-2}$ eV. The parameter, $\Lambda$, basically a Clebsch-Gordan coefficient with the value of about 0.4, is NOT an expansion parameter, in contrast with the Wolfenstein parameter, $0.22 < \lambda < 0.25$ needed to explain quark masses.

The phase, δ, of the mixing matrix U has a serious effect on the mass matrix only for the matrix elements $M_{e\mu}$ and $M_{e\tau}$, because these are the only ones for which the real part vanishes in the allowed range for $\theta_{13}$. Their dependence on $s_{13}$ for various values of δ is given explicitly.

A study of the elements of the mass matrix, M, for our solution 1, that of the perfect rational hierarchy, for the case $\delta = 0$, shows that all of them are smaller than 0.03 eV. The only candidates for double texture zero models are $M_{ee}$ and $M_{e\mu} = M_{\mu e}$ (or $M_{e\tau} = M_{\tau e}$, with $\theta_{13} \to -\theta_{13}$).

**ACKNOWLEDGEMENTS**

We thank Dr. Douglas Michael for a useful discussion of neutrino data. We also are grateful for the support of the Caltech High Energy Theory Group and the U. S. Department of Energy in the preparation of this manuscript. We thank the Aspen Center for Physics for its kind hospitality.